\documentclass[a4paper,twoside]{article}
\newcommand{\affil}[1]{$^{\rm #1}$}
\usepackage[authoryear]{natbib}
\bibpunct{(}{)}{;}{a}{}{,}
\usepackage{graphicx}
\date{} 
\newcommand{\mbf}[1]{\mbox{\boldmath $#1$}}

\newcommand{\Eqn}[1]{Equation~(\ref{eqn:#1})}

\newcommand{\eqn}[1]{equation~(\ref{eqn:#1})}

\newcommand{\Sec}[1]{Section~\ref{sec:#1}}

\newcommand{\Fig}[1]{Figure~\ref{fig:#1}}
\newcommand{\Figs}[2]{Figures~\ref{fig:#1} and~\ref{fig:#2}}

\newcommand{\Offt}{\ensuremath{ {\mathcal O}_{\mathrm{FFT}} }}
\newcommand{\flops}{\ensuremath{ \mathrm{flops} }}

\newcommand{\Ci}{\ensuremath{i}}

\newcommand{\trace}{{\rm tr}}

\newcommand{\pauli}[1]{\ensuremath{ {\mbf{\sigma}}_{#1} }}

\newcommand{\polyco}{\ensuremath{ {\mathcal P}_{\phi} }}

\newcommand{\dspsr}{{\sc dspsr}}
\newcommand{\psrchive}{{\sc psrchive}}
\newcommand{\psrfits}{{\sc psrfits}}

\newcommand{\intel}{Intel$^{\scriptsize\textregistered}$}
\newcommand{\xeon}{Xeon$^{\scriptsize\textregistered}$}

\newcommand{\nvidia}{NVIDIA$^{\scriptsize\textregistered}$}
\newcommand{\cuda}{CUDA} 

\newcommand{\nfft}{\ensuremath{ N_{\mathrm{FFT}} }}

\newenvironment{squishlist}{
\begin{list}{$\bullet$\ }
 {\setlength{\itemsep}{0pt}
     \setlength{\parsep}{0pt}
     \setlength{\topsep}{0pt}
     \setlength{\partopsep}{0pt}
     \setlength{\leftmargin}{0em}
     \setlength{\labelwidth}{0em}
     \setlength{\labelsep}{0em}}
}{\end{list}}

\title{\large\bf\flushleft 
\dspsr: Digital Signal Processing Software for Pulsar Astronomy }

\author{\parbox{\textwidth}{\flushleft
\vspace{-0.5cm}
{\it W.\ van Straten\affil{A,B}, and
     M.\ Bailes\affil{A}}\\
\vspace{0.4cm}
{\small \affil{A}\,Swinburne University of Technology, PO Box 218, Hawthorn VIC 3122, Australia}\\
{\small \affil{B}\,Email: vanstraten.willem@gmail.com}}}

\begin{document}
\twocolumn[
\begin{changemargin}{.8cm}{.5cm}
\begin{minipage}{.9\textwidth}
\vspace{-1cm}
\maketitle
%
%
\small{\bf Abstract:}

\dspsr\ is a high-performance, open-source, object-oriented, digital 
signal processing software library and application suite for use in
radio pulsar astronomy.
Written primarily in C++, the library implements an extensive range of
modular algorithms that can optionally exploit both multiple-core
processors and general-purpose graphics processing units.
After over a decade of research and development, \dspsr\ is now stable
and in widespread use in the community.
This paper presents a detailed description of its functionality,
justification of major design decisions, analysis of phase-coherent
dispersion removal algorithms, and demonstration of performance on
some contemporary microprocessor architectures.




\medskip
\medskip
\end{minipage}
\end{changemargin}
]
\small


\section{Introduction}

The main technical challenges that present themselves to pulsar
astronomers arise from two opposing forces: the drive for
greater sensitivity and the adverse effects of propagation through the
interstellar medium (ISM).  For a given antenna and observing
schedule, greater sensitivity may be achieved by increased
instrumental bandwidth, which comes at the cost of additional signal
distortion primarily due to plasma dispersion in the ISM.  A wide
variety of pulsar instrumentation has been developed to either
mitigate or eliminate interstellar dispersion; these may be broadly
classified as either post-detection or pre-detection techniques.

Post-detection methods employ a filterbank, either analog or digital,
to divide the wideband signal into narrow channels. The voltage signal
in each frequency channel is detected and the inter-channel dispersion
delays are removed before integrating over frequency. Post-detection
dispersion removal is fundamentally limited by residual intra-channel
dispersion smearing and reduced temporal resolution, both of which are
inversely proportional to the number of filterbank channels used.
In contrast, pre-detection (or phase-coherent) dispersion removal,
completely eliminates dispersion smearing while retaining the original
time resolution of the observed signal \citep{hr75}.
%
%
Phase-coherent dedispersion is applied directly to the voltage signal,
which must be sampled at the Nyquist rate and either recorded or
reduced using a high-speed digital signal processing system.
Owing to the relatively high demand on both data storage and
computing resources, the earliest baseband recording and
processing systems were limited to relatively small bandwidths.
For example, coherent dedispersion was pioneered using a baseband
recorder with a modest bandwidth of 125\,kHz, which could record with
a 20\% duty cycle for only 3 minutes before filling the magnetic tape
\citep{han71}.
%

Fortunately, the capacity of baseband recording and processing systems
closely follows the continual growth of affordable, commercial
computing and data storage technologies,
and by the mid-1990s a number of baseband recorders were in regular use
at radio observatories around the world.
The Princeton Mark IV system, capable of continuously 2-bit sampling
a $2\times10$\,MHz bandpass \citep{shr97,sta98,sst+00}, was in use at the
Arecibo and Jodrell Bank observatories.
At Parkes, baseband recorders included
the Wide Bandwidth Digital Recording (WBDR) system, capable of
recording a $2\times50$\,MHz bandpass \citep{jcpu97};
the S2 VLBI recorder, a $2\times16$\,MHz system \citep{wvd+98};
and the Caltech Parkes Swinburne Recorder \citep[CPSR;][]{vbb00}, a
$2\times20$\,MHz version of the Caltech Baseband Recorder (CBR) at Arecibo.
Each of the above systems recorded the baseband signal to magnetic
tape for transport and offline analysis on high-performance computing
resources.
Also in use at Green Bank, Effelsberg, and Arecibo were versions of
the coherent Berkeley Pulsar Processor (cBPP), which employed digital
signal processing hardware to perform coherent dedispersion in real
time.  Depending on the pulsar dispersion measure, these systems were
capable of correcting a bandpass of up to $2\times112$\,MHz
\citep{bdz+97}.

At Parkes, CPSR2 led the evolution away from recording on
magnetic tape by using a modest cluster of workstations at the
telescope to process $2\times128$\,MHz of bandwidth in quasi-real time
\citep{bai03}.  The Arecibo Signal Processor (ASP), Green Bank
Astronomical Signal Processor (GASP), and Berkeley Orleans Nancay
processor (BON) series of instruments also process $2\times128$\,MHz in
quasi-real time, using 8-bit sampling to significantly reduce
quantisation distortions \citep{drb+04a}.  The second generation of
the Dutch Pulsar Machine, PuMa-II, another high dynamic range system
with 8-bit sampling and $2\times160$\,MHz of bandwidth is now in operation at
the Westerbork Synthesis Radio Telescope \citep{ksv08}.  Each of the
above systems use large capacity, high speed disks to buffer the
incoming baseband signal before it is processed.

In 2007, the first coherent dedispersion software to work completely
in memory was pioneered at Parkes with the development of the ATNF
Parkes Swinburne Recorder (APSR).
This instrument is capable of real-time processing $2\times1$\,GHz of
bandwidth with 2 bits/sample, or $2\times256$\,MHz with 8 bits/sample.
APSR uses the same digital signal processing software that was
developed at Swinburne to reduce the baseband data recorded by the S2
and CPSR instruments.
Later named \dspsr\footnote{http://dspsr.sourceforge.net}, 
this software evolved through the development of PuMa-II at WSRT, and
is now utilised to process data from a wide variety of instruments,
including the Australian Long Baseline Array Data Recorder (LBADR) and
the Giant Metrewave Radio Telescope (GMRT) software backend \citep{rgp+09}.
The real-time processing requirements of APSR motivated the
implementation of the shared memory and multi-threaded capabilities of
\dspsr.  

Around this time, advances in general-purpose computing on graphics
processing units (GPUs) were rapidly transforming the state of the art
of digital signal processing.
The Green Bank Ultimate Pulsar Processing Instrument (GUPPI) first
demonstrated the use of GPUs to perform phase-coherent dispersion
removal in real time (P.~Demorest 2009, private communication).
Following these remarkable developments, \dspsr\ was extended to use
the Compute Unified Device Architecture (\cuda) developed by
\nvidia\footnote{http://www.nvidia.com/cuda}.
This new capability was driven by the development of a GPU-based
digital signal processing system for Parkes using the Interconnect
Break-out Board (IBOB) produced by the Center for Astronomy Signal
Processing and Electronics Research (CASPER) at Berkeley.  The CASPER
Parkes Swinburne Recorder (CASPSR) performs real-time phase-coherent
dispersion removal on $2\times400$\,MHz using four server-class
workstations, each equipped with two \nvidia\ Tesla C1060 GPUs.

Through these stages of evolution and refinement, \dspsr\ has amassed
an extensive range of useful features.  These are briefly summarised in
\Sec{algorithms}, followed by a more detailed description of the 
algorithms developed for radio pulsar astronomy.
The performance of the library is demonstrated in \Sec{performance}
using currently available technology and common observing
configurations.
In \Sec{conclusion}, the future of the software project is discussed
with regard to the upcoming generation of radio telescopes.


\section{Algorithms and Features}
\label{sec:algorithms}

The \dspsr\ software processes continuous streams of radio pulsar
astronomical data, producing integrated statistics such the
phase-resolved average polarisation of the pulsar signal. The salient features of this software are
\begin{squishlist}
\item automatic excision of invalid data, such as those lost during
      transfer or corrupted by impulsive interference;
\item correction of digitisation distortions via
      dynamic output level setting \citep{ja98};

\item phase-coherent dispersion removal \citep{hr75} with optional
      Jones matrix convolution for high time-resolution polarimetry
      \citep{van02};

\item synthetic filterbank formation with concurrent coherent 
      dedispersion and/or Jones matrix convolution;

\item full-polarimetric detection;

\item computation of phase-resolved averages (folding) using
      either a polynomial approximation to the pulsar phase model, a
      constant period, or acceleration search parameters;

\item formation of sub-integrations of arbitrary length, including
      single pulses;

\item simultaneous folding of multiple pulsars, such as globular cluster 
      or double pulsars;

\item exploitation of multiple processing cores using parallel
      computing threads;

\item accelerated computation on graphics processing units;

\item transparent buffering of input data to handle the edge
      effects of certain algorithms, such as the overlap-save method
      of discrete convolution;

\item real-time signal processing directly from a ring buffer in
      shared memory; and

\item time-division multiplexing: discontiguous segments of data may
      be processed as a single stream to minimise initialisation time.

\end{squishlist}

The following sections describe these features in greater detail while
outlining a typical pulsar signal processing path.

\subsection{Invalid Data Excision}
\label{sec:bad}

Any operation may flag sections of data as invalid, such that they are
ignored by subsequent components in the signal processing chain.  This
design feature is implemented as an array of weights that is
maintained in parallel with the blocks of data that represent the
astronomical signal.  The weights array is typically used to flag
segments of data in the time domain, such that a single weight applies
to all frequency channels and polarisations over some epoch.  However,
the weights array can also be used to flag one or more specific
frequency channels as invalid.

\begin{figure}
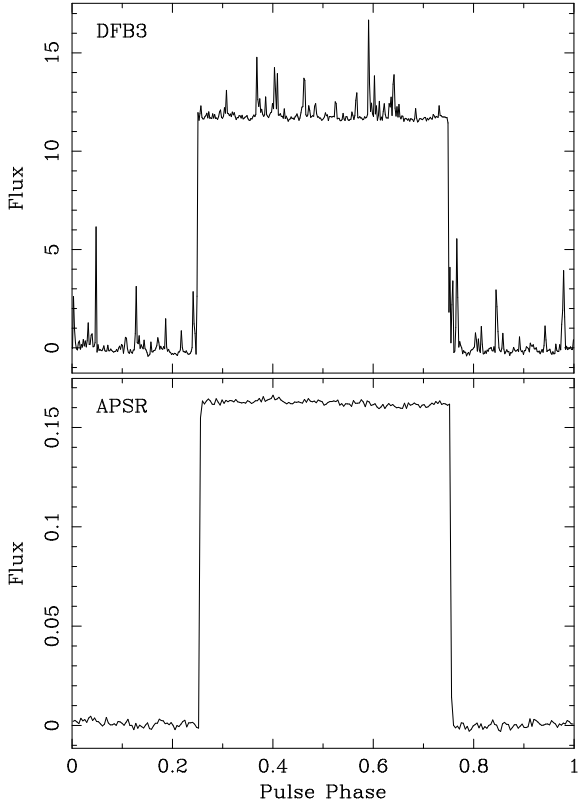

\includegraphics[angle=-90,width=74.6mm]{DFB3_cal.eps}
\vspace{5mm}
\includegraphics[angle=-90,width=75mm]{APSR_cal.eps}
\caption
{Automatic impulsive interference excision by \dspsr\ during a
lightning storm.  A noise diode coupled to the receptors of the Parkes
20\,cm Multibeam receiver was modulated by a square wave with a 50\%
duty cycle.  This calibrator signal was observed at a centre frequency
of 1369\,MHz using the Parkes Digital Filterbank (PDFB3, above) and
the ATNF Parkes Swinburne Recorder (APSR, below).  Both instruments
observed the same $2\times256$\,MHz band and the flux scale in each
plot is uncalibrated.  Each of the four 30-second integrations output
by PDFB3 are irreversibly corrupted by broadband impulsive noise.
However, by discarding only the contaminated segments of APSR data
(less than 0.05\% of the signal) the 2-minute integration produced by
\dspsr\ is greatly improved.}
\label{fig:lightning}
\end{figure}

Many of the classes used to convert between $n$-bit and floating point
representations of the digitised signal monitor the data and flag
sections as invalid whenever certain statistics (such as the noise
power) fall outside the acceptable ranges of operation.  The weights
array is used by operations that perform integrations (such as the
pulse profile folding operation described in \Sec{folding}) to ensure
that final results are not corrupted by invalid data.  For example,
invalid data may be recorded before the instrumentation is properly
initialised.  As illustrated in \Fig{lightning}, strong bursts of
impulsive interference such as lightning are completely excised using
this technique.



\subsection{Dynamic Output Level Setting}
\label{sec:dynamic_level}

Analog-to-digital conversion is a non-linear process that introduces
both noise and signal distortion.  Dynamic output level setting aims
to correct the quantisation distortion by restoring the linear
relationship between digitised and undigitised power
\citep[][hereafter JA98]{ja98}.  In \dspsr\ this correction is
performed during the conversion from 2-bit to floating point
representations of the digitised voltage samples.  Referring to
section 6 of JA98, the data are divided into consecutive segments of
$L$ digitised samples.  In each segment, the number of low voltage
states, $M$, is counted and the undigitised power, $\sigma^2$, is
estimated by inverting equation 45 of JA98,

\begin{equation} 
\Phi={M\over L}={\rm erf}\left(\frac{t}{{\sqrt2}\sigma}\right), 
\label{eqn:psierf}
\end{equation}
where the error function, 
\begin{equation}
{\rm erf}(x)=\frac{2}{\sqrt\pi}\int_0^x e^{-y^2}dy,
\end{equation} 
and $t$ is the optimal threshold between low and high voltage
states. \Eqn{psierf} is solved for $\sigma$ using the Newton-Raphson
method, and it is computationally prohibitive to perform this
iterative calculation for every $L$-point segment.  
However, as there are only $L+1$
possible values of $M$, \dspsr\ utilises a lookup table of
precomputed output levels, which are stored for only an acceptable
range of input power estimates.  Those segments of data for which $M$
falls outside of the acceptable range are flagged as invalid (as
described in more detail the next section).  \dspsr\ also
maintains a histogram of occurrences of $M$ that is archived for later
use as a diagnostic tool when assessing the quality of the baseband
recording.

\subsection{Coherent Dedispersion}
\label{sec:coherent_dedispersion}

Electromagnetic waves propagating through the interstellar medium
(ISM) experience phase dispersion that effects a frequency-dependent
group velocity.  Describing the ISM as a cold, tenuous plasma, the
frequency response function, 
\begin{equation}
H(\nu+\nu_0)=\exp\left(\Ci{2\pi D\nu^2\over\nu_0^2(\nu+\nu_0)}\right),
\label{eqn:dispersion_frequency_response}
\end{equation}
is obtained and used to deconvolve the observed radio signal
\citep{han71,hr75}.  Here, $\nu_0$ is the centre frequency of the
observation and the dispersion $D$ is related to the more commonly
used dispersion measure $DM$ by \citep{mt72}
\begin{equation}
\label{eqn:dm}
DM\,({\rm pc\,cm^{-3}})=2.41\times 10^{-4}D\,({\rm s\,MHz^{2}}).
\end{equation}
The dispersion measure is equal to the integral of the free electron
density along the line of sight to the pulsar.  Although the constant
of proportionality has been derived with greater precision \citep{bhvf93},
%
%
\Eqn{dm} is the standard adopted by pulsar astronomers and implemented
by \dspsr.

The duration of the phase dispersion impulse response
function is called the sweep or smearing time, $t_d$.  It is
equivalent to the width of a band-limited delta-function after passage
through the ISM, and is a function of the bandwidth, $\delta\nu$, and
centre frequency, $\nu_0$, such that
\begin{equation}
\label{eqn:sweep_time}
	t_d=D(\nu_{\rm min}^{-2}-\nu_{\rm max}^{-2}),
\end{equation}
where $\nu_{\rm min}=\nu_0-\delta\nu/2$, and $\nu_{\rm
max}=\nu_0+\delta\nu/2$.  

\begin{figure}
\includegraphics[angle=-90,width=70mm]{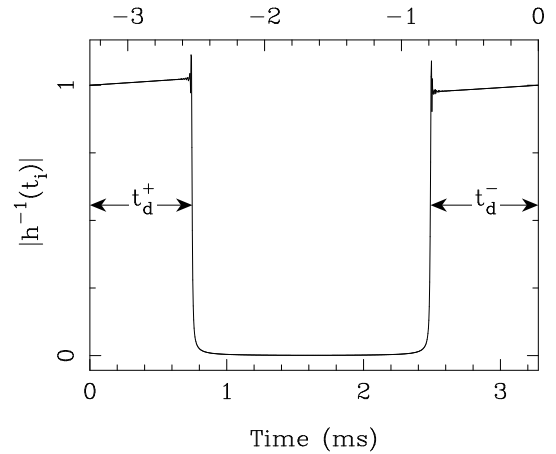}
\caption
{The magnitude of the dedispersion impulse response function,
$|h^{-1}(t_i)|$, as calculated from the inverse FFT of the frequency
response function, $H^{-1}(\nu_k)$, and normalized to unity at the
origin ($t=0$).  For this example, $\nu_0$=600\,MHz,
$\delta\nu$=20\,MHz, $DM$=2, and $N$=65536 points.
%
%
Reflecting the cyclicity of discrete convolution, time is labelled in
both the positive (bottom axis) and negative (top axis) directions
with respect to the origin, which recurs at the left and right
boundaries of the plot.
%
%
At the edges of the response function, where the magnitude drops to zero,
the ringing is due to the cyclic discontinuity in $H^{-1}(\nu_k)$ at $\nu=0$.
The smearing in the upper half of the band, $t_d^+=750\mu$s, and lower
half of the band, $t_d^-=788\mu$s, result in asymmetry about the origin.
%
%
This asymmetry also causes the observed slope in the magnitude of the
impulse response; the flux at low frequency is smeared over more time
than the flux at high frequency, but the integrated fluxes in the two
halves of the band are equal.
Note that $t_d^+$ and $t_d^-$ correspond to $m_+$ and $m_-$ in Figure
13.1.3 of Numerical Recipes.}
\label{fig:impulse_response}
\end{figure}

\begin{figure*}
\centerline{\includegraphics[angle=0,width=11cm]{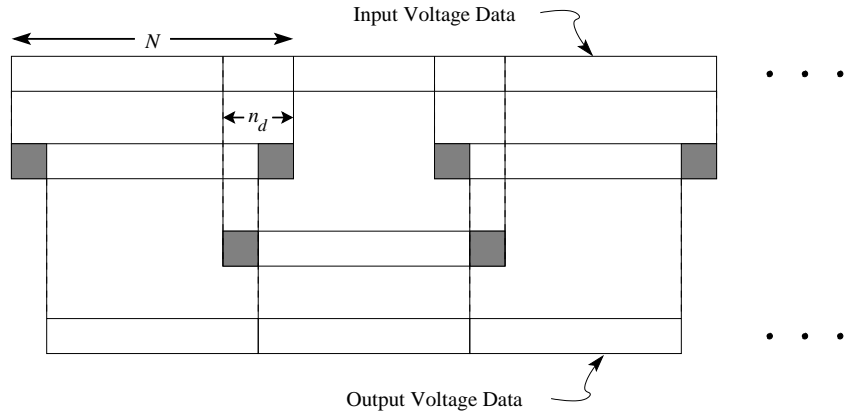}}
\caption
{The overlap-save method of discrete convolution. 
The input data (top) are divided into $N$-point segments that overlap
by the number of non-zero points, $n_d$, in the discrete phase
dispersion impulse response function.  Each $N$-point segment is
separately transformed into the Fourier domain, multiplied by the
frequency response function, $H^{-1}(\nu_k)$, and transformed back
into the time domain (middle).  Polluted points from the wrap-around
region (shaded grey) are discarded before copying the result into the
output data stream (bottom). }
\label{fig:overlap_save}
\end{figure*}

Deconvolution is most efficiently performed in the frequency domain,
where the observed signal is simply multiplied by the inverse of the
discrete form of \Eqn{dispersion_frequency_response}.  This results in
cyclical convolution, as described in Chapter 18 of \citet{bra86b} and
Section~13.1 of Numerical Recipes \citep[][hereafter NR]{ptvf92}.
\Fig{impulse_response} (cf.\ Figure 13.1.3 of NR) illustrates that each 
time sample output by the cyclical convolution operation will depend
upon the $n_d^+=rt_d^+$ points preceding it and the $n_d^-=rt_d^-$
points following it, where $r$ is the sampling rate and $t_d^+$ and
$t_d^-$ are the dispersion smearing times in the upper and lower halves
of the band. 
Owing to the periodicity assumption of the discrete Fourier transform,
the first $n_d^+$ points in the result of the convolution operation
will be erroneously mixed with data wrapped around from the end of the
data segment, and the last $n_d^-$ points of the result will be
erroneously mixed with time samples from the beginning of the input
segment \citep[cf.\ Figure 18.4 of][]{bra86b}. 
Consequently, the polluted points (called the wrap-around region) from
the result of each transformation step are discarded, a technique that
forms part of the overlap-save method of discrete convolution.  
This method, depicted in \Fig{overlap_save}, also accounts for the
fact that the duration of the impulse response function is generally
much shorter than that of the signal to be convolved.

Noting the quadratic form of \Eqn{sweep_time}, it can be seen that the
lowest frequency must be advanced by an amount greater than the delay
applied to the highest frequency.  This results in asymmetry of
$h(t_i)$ around $t=0$, which must be accounted when discarding
wrap-around points from the result of each cyclical convolution
operation.
As $n_d=n_d^-+n_d^+$ points are discarded, the number of samples $N$
in the result of the Fast Fourier Transform (FFT) must be greater than
$n_d$.  $N$ is chosen by minimising the number of floating point
operations per second,
\begin{equation}
\flops={\Offt(N)\over N-n_d} 2 \delta\nu,
\label{eqn:flops}
\end{equation}
where \Offt($N$) is the number of floating point operations required
to compute the FFT of $N$ complex input samples, which asymptotically
approaches $5N\log_2N$ when $N$ is an integer power of
two\footnote{http://www.fftw.org/speed}.
This equation is valid for both real-valued and complex-valued input
data (see the Appendix for discussion).  The factor of 2 arises
because both forward and backward transforms are computed.
\Eqn{flops} provides an estimate of the minimum number of \flops\ 
required to process the data in real time.  This number can be
directly compared with FFT benchmarks, such as those computed by
{\sc benchFFT}\footnote{http://www.fftw.org/benchfft}.  As
discussed in more detail in \Sec{performance}, actual performance
measurements can be used to better select the optimal FFT length.

To perform the FFT, \dspsr\ can utilise both open source and
proprietary FFT libraries, including
the Fastest Fourier Transform in the West
\citep[FFTW;][]{fj05}
the \intel\ Math Kernel Library\footnote{http://www.intel.com/software/products/mkl}
and \intel\ Integrated Performance Primitives 
(IPP)\footnote{http://www.intel.com/software/products/ipp}.
On allocation, all arrays are aligned on 16-byte boundaries so that
vector instruction sets, such as Streaming SIMD Extensions (SSE), may
be exploited.

{\bf Invalid Data Excision}:
Referring to \Fig{overlap_save}, if any of the input voltage data in
an $N$-point segment are flagged as invalid, then the all of the
$N-n_d$ samples of the output voltage data for that cyclical
convolution step will be flagged.

\subsection{Synthetic Filterbank}
\label{sec:filterbank}

Though phase-coherent dispersion removal retains the original temporal
resolution of the digitised signal, this resolution is often
integrated away when forming the average pulse profile.  It therefore
proves useful to trade some time resolution for frequency resolution
by dividing the observed band into a number of sub-bands, or channels.
Both the smearing time, $t_d$, and the sampling rate, $r$, in each
sub-band is reduced, resulting in smaller $n_d$; this reduces the
transform length, $N$, required for subsequent coherent dedispersion,
and thereby improves computational efficiency.  Furthermore, with
greater frequency resolution the measured polarisation may be better
calibrated, or individual sub-bands corrupted by narrow-band
radio-frequency interference (RFI) may be deleted.  Finally, when
frequency resolution is retained, a better estimate of the dispersion
measure may be applied at a later time to correct the inter-channel
dispersion delays before further integrating in frequency.
Using \dspsr, the observed signal may be divided into $N_c$
frequency channels using either the deprecated or convolving filterbank,
both of which are based on the FFT.

\subsubsection{Deprecated Filterbank}
\label{sec:deprecated_filterbank}

In the deprecated filterbank, the data are divided into
non-overlapping segments of $N_c$ complex points (or 2$N_c$ real
points).  Each segment is transformed into the Fourier domain, where
each of the $N_c$ complex spectral values is treated as a time sample
in one of $N_c$ independent signals, each with bandwidth,
$\delta\nu^\prime=\delta\nu/N_c$, and sampling rate, $r^\prime=r/N_c$.
Phase-coherent dispersion removal is then separately performed on each
of the $N_c$ resulting channels using unique phase dispersion
frequency response functions, each tuned to the centre frequencies of
the output filterbank channels.

\begin{figure}
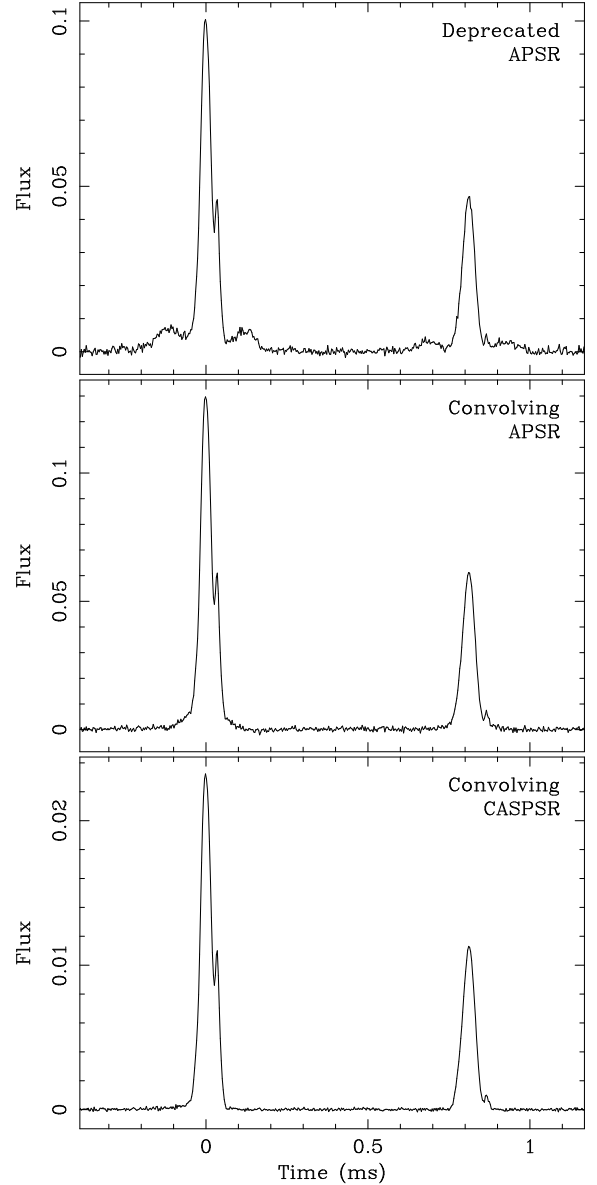

\includegraphics[angle=-90,width=75mm]{deprecated.eps}
\includegraphics[angle=-90,width=75mm]{convolving.eps}
\includegraphics[angle=-90,width=75mm]{caspsr.eps}
\caption
{Correction of spectral leakage using the convolving filterbank.
PSR\,B1937+21 was observed at a centre frequency of 1369\,MHz using
the $2\times256$\,MHz band of APSR (top two panels) and at a centre
frequency of 1382\,MHz using the $2\times400$\,MHz band of CASPSR
(bottom panel).  The APSR data were processed using a synthetic
filterbank as implemented by the deprecated (top panel) and convolving
(middle panel) methods.  The residual distortion in the APSR data
(most obvious is the triangular base of the main pulse in the middle
panel) is due to the Parkes Digital Filterbank, as described in the
text.  The uncorrupted CASPSR observation best demonstrates the
fidelity of the convolving filterbank.}
\label{fig:spectral_leakage}
\end{figure}

This method of synthetic filterbank formation is equivalent to the
coherent filterbank described in Section~3.1.3 of \citet{jcpu97}.
Although it reduces the computational cost of coherent dedispersion
(see the following section and the Appendix for further discussion),
the technique suffers from the spectral leakage of the discrete
Fourier transform, which mixes a significant amount of power between
neighbouring frequency channels (NR, Section~13.4).  When the
inter-channel dispersion delay is large, the artifacts of spectral
leakage are readily observed as delayed images of the pulse profile,
as shown in \Fig{spectral_leakage}.  

Here, the dispersion delay between the 500\,kHz channels of the
deprecated filterbank (which ranges from $\sim$ 88 to 154\ $\mu$s
across the band) is seen between the peaks (main and inter-pulse) and
the leakage artifacts on either side of them.  The remaining artifacts
in the APSR convolving filterbank are introduced by the Parkes Digital
Filterbank (PDFB3), which is used to digitise and divide the 256\,MHz
band into 16 smaller 16\,MHz bands.  The two-stage (1024-channel
analysis followed by $16\times 64$-channel synthesis) polyphase
filterbank implemented by the PDFB3 introduces spectral leakage
between the 250\,kHz channels produced by the analysis filterbank; the
dispersion delay between these channels ranges from $\sim$ 44 to
77\,$\mu$s across the band.  Work is currently underway to correct
this artifact using a single-stage 16-channel analysis polyphase
filterbank.  

In contrast, CASPSR uses time-division demultiplexing to implement
parallel processing on multiple GPUs; therefore, spectral leakage is
limited to that introduced by the forward 512\,k-point FFT in the
convolving filterbank used to process these data.  At this frequency
resolution, the dispersion delay between neighbouring channels varies
between only $\sim$ 0.11 and 0.27 $\mu$s, which is less than the
temporal resolution of the folded profile.

\subsubsection{Convolving Filterbank}
\label{sec:convolving_filterbank}

The spectral leakage function may be improved by the application of a
window function in the time domain (NR, Fig.\ 13.4.2).  However, the
physical bandwidth of the spectral leakage function may also be
decreased by increasing the length of the Fourier transform by a
factor, $N^\prime$.  Where $K=N^\prime\times N_c$, each $K$-point
segment of data is transformed into the Fourier domain, and the result
is divided into $N_c$ non-overlapping segments of $N^\prime$ spectral
points.  Each segment is inverse transformed back into the time
domain, producing $N^\prime$ time samples from $N_c$ independent
filterbank channels.

Coherent dedispersion may be performed simultaneously with this
filterbank technique.  While in the Fourier domain, each of
the $N_c$ spectral segments is multiplied by a unique dedispersion
frequency response function.  In this case, $N^\prime$ must be
chosen to match the size of the response function.  As the same
inverse transform size is applied to each sub-band, $N^\prime$ must be large
enough to accommodate the maximum smearing time,
$t_d^\prime=D(\nu_{\rm min}^{-2}-\nu_{\rm max}^{\prime\,-2})$,
corresponding to the sub-band with the lowest centre frequency, where
$\nu_{\rm min}=\nu_0-\delta\nu/2$, and $\nu^\prime_{\rm max}=\nu_{\rm
min}+\delta\nu/N_c$.  Owing to the quadratic form of \Eqn{sweep_time},
$t_d^\prime >~ t_d/N_c$.  Although the smearing is less severe in
those sub-bands with higher centre frequencies, symmetry dictates
that $n_d^\prime=r^\prime t_d^\prime >~ n_d / N_c^2$ wrap-around
points are discarded from the result of each of the $N_c$ inverse
transforms (cyclical convolution products).  Each $K$-point segment of
input data therefore overlaps by $N_c\times n_d^\prime$ points.

As shown in the Appendix, phase-coherent dispersion removal using
either the deprecated or convolving filterbank methods requires the same
number of floating point operations per second,
\begin{equation}
\flops^\prime =
{ 5 N^\prime (\log_2N_c + 2\log_2N^\prime) 
  \over
  N^\prime-n_d^\prime } \delta\nu.
\label{eqn:filterbank_flops}
\end{equation}
Note that the size of the first FFT, $K=N_c\times N^\prime > n_d/N_c$, is
inversely proportional to the number of filterbank channels, $N_c$.
As smaller transform lengths improve processing efficiency, $N_c$
should be chosen to be as large as possible without sacrificing
required time resolution.  This remains true only up to the limit
where the resulting sampling interval, $t_s^\prime=1/r^\prime$, becomes
larger than $t_d^\prime$ (i.e. where $n_d^\prime\le1$).

As shown in \Fig{spectral_leakage}, the convolving filterbank
method eliminates the spectral leakage artifacts produced using the
deprecated method.  Synthetic filterbank formation with
simultaneous convolution is summarised by the
following algorithm:
\begin{enumerate}
\item Divide the signal into $K$-point segments 
	($K=N_c\times N^\prime$) that overlap by $N_c\times n_d^\prime$
	points (cf. top of \Fig{overlap_save}).
\item For each segment, perform the $K$-point FFT, and divide
	the result into $N_c$ non-overlapping segments, or sub-bands,
	of $N^\prime$ points.
\item For each of the $N_c$ sub-bands:
\begin{enumerate}
\item multiply the $N^\prime$-point sub-band by its unique frequency 
	response function;
\item perform an $N^\prime$-point inverse FFT;
\item discard the $n_d^\prime$ wrap-around points; and
\item copy the remaining time samples into the corresponding 
      output frequency channel.
\end{enumerate}
\end{enumerate}

{\bf Invalid Data Excision}:
In addition to the discussion of invalid data excision in
\Sec{coherent_dedispersion}, synthetic filterbank formation reduces
the number of time samples per weight by a factor of $N_c$.  If $N_c$
is greater than the number of samples per weight, then the weights
array will be rebinned with care to ensure that invalid data remain
flagged.


\subsection{Detection}
\label{sec:integrate}

If the voltage signals from both receptors of a dual-polarisation
receiver are available, \dspsr\ computes the polarisation of the
electromagnetic radiation.
Radio pulsar polarimetry yields information about the physics of the
emission mechanism \citep[e.g.][]{es04}, the geometry of the pulsar's
magnetosphere \cite[e.g.][]{jhv+05}, and the average magnetic field of
the ISM along the line of sight \cite[e.g.][]{hml+06}.
It also provides additional constraints that may be
exploited for high-precision pulsar timing \citep{van06}.

Polarisation is described by the second-order statistics
of the transverse electric field vector $\mbf{e}$, as given by
the complex $2\times2$ coherency matrix,
$\mbf{\rho}\equiv\langle\mbf{e\, e}^\dagger\rangle$ \citep{bw80}.
Here, the angular brackets denote an ensemble average,
$\mbf{e}^\dagger$ is the Hermitian transpose of $\mbf{e}$, and an
outer product is implied by the treatment of $\mbf{e}$ as a column
vector.  
A pulse phase-resolved ensemble average is performed by the folding
operation described in the next section. When detecting synthetic
filterbank data, the coherency matrix is computed independently in
each frequency channel.

A useful geometric relationship between the complex two-dimensional
space of the coherency matrix and the real four-dimensional space of
the Stokes parameters is expressed by the following pair of equations:
\begin{eqnarray}
{\mbf\rho} & = & S_k\,\pauli{k} / 2     \label{eqn:combination} \\
S_k & = & \trace(\pauli{k}\mbf{\rho}).  \label{eqn:projection}
\end{eqnarray}
Here, $S_k$ are the four Stokes parameters, Einstein notation is used
to imply a sum over repeated indeces, $0\le k \le3$, $\pauli{0}$ is
the $2\times2$ identity matrix, $\pauli{1-3}$ are the Pauli matrices,
and $\trace$ is the matrix trace operator.  The Stokes four-vector is
composed of the total intensity $S_0$ and the polarisation vector,
$\mbf{S}=(S_1,S_2,S_3)$.  \Eqn{combination} expresses the coherency
matrix as a linear combination of Hermitian basis matrices;
\eqn{projection} represents the Stokes parameters as the projections
of the coherency matrix onto the basis matrices.  


The fourth-order moments of the electric field, as described by
four-dimensional covariance matrix of the Stokes parameters, can
optionally be formed and averaged as a function of pulse phase.  The
covariances of the Stokes parameters provide additional information
about the pulsar radiation, such as the presence of orthogonally
polarised modes of emission and the degree of correlation between
mode intensities \citep{ms98,van09}.


\subsection{Folding}
\label{sec:folding}

Ensemble-averages of the detected signal are computed as a function of
topocentric pulse phase during a process commonly called folding.
Apparent topocentric pulse phase is determined using a polynomial
approximation to the pulsar timing model, $\polyco(t)$, generated
using either the
{\sc tempo}
\citep{tw89}\footnote{http://pulsar.princeton.edu/tempo} or
{\sc tempo2} 
\citep{hem06}\footnote{http://www.atnf.csiro.au/research/pulsar/tempo2}
software packages.  
The detected signal of interest, $Y(t_i)$, is folded by integrating each
of its samples into one of $n$ equally spaced phase bins, $Y(\phi_k)$,
where bin number, $k$, is given by 
\begin{equation}
k=(n\polyco(t_i))\,{\rm mod}\,n,
\end{equation} 
where \polyco\ has dimensionless phase (turns) and the mod operator
returns the remainder after division, such that $0\le k<n$.

Typically, a vector of processes, $\mbf{Y}(t_i)$ is folded in parallel
with one count, $N(\phi_k)$, of the number of time-samples accumulated
in each phase bin.
For example, $\mbf{Y}(t_i)$ may represent the four Stokes parameters
in each of the $N_c$ frequency channels.
The average pulse profile of each process is then given by
\begin{equation}
\langle Y_j(\phi_k)\rangle=Y_j(\phi_k)/N(\phi_k).
\end{equation}
The start and end times, $t_0$ and $t_N$, of the observation are
used to assign the mean time of the rising edge of phase bin zero.
Under the assumption that \polyco\ and its inverse may
be calculated with sufficiently high accuracy, this is given by
\begin{equation}
\tau_0 = \polyco^{-1} ( \polyco( (t_0+t_N)/2 )\,{\rm mod}\,1)
\label{eqn:tau_0}
\end{equation}

During the folding operation, \dspsr\ can divide the signal into
segments with arbitrary start and end times and produce a unique set
of pulse profiles for each segment.  This feature can be used to
divide a long observation into regular intervals, or sub-integrations,
that may be later combined using an updated version of the pulsar
timing model, or following the deletion of sub-integrations that have
been corrupted by instrumentation faults or radio frequency
interference.

The start and end times of each sub-integration can also be made to
coincide (within the temporal resolution of the signal) with integer
values of pulse phase, thereby producing single-pulse profiles.
In single-pulse mode, it is usually necessary to remove the
inter-channel dispersion delays, such that each single pulse integration
output by \dspsr\ contains the same pulse in all frequency channels.
It may also be necessary to adjust the reference phase of phase bin
zero so that the on-pulse region of pulse phase is not divided across
multiple integrations.

For pulsars with short spin periods, such as millisecond pulsars, it
may not be feasible to store every single pulse with full frequency
resolution.  In such cases, it becomes desirable to either integrate
the data down to some manageable size or add some conditions such that
only a subset of the single pulse integrations are kept.  
To this end, \dspsr\ employs the \psrchive\ \citep{hvm04} command language
interpreter\footnote{http://psrchive.sourceforge.net/manuals/psrsh} to
run a user-supplied script, through which a diverse range of
post-processing tasks can be executed.
For example, when studying the polarisation of single pulses, the
\psrchive\ script can be used to calibrate before integrating over
frequency channels.
When searching for giant pulses, the \psrchive\ script can test if the
single pulse contains any signal that is a certain threshold
above the noise.

Multiple pulsars can be folded simultaneously, which is particularly
useful when observing globular cluster or double pulsar systems.
When folding multiple pulsars, the end of the signal path is forked
across multiple instances of the folding operation.  That is, the
signal is processed identically for each pulsar up to the final stage
of folding.
Therefore, when performing phase-coherent dispersion removal on a
set of globular cluster pulsars, the mean or median dispersion 
measure should be used.

When parallel threads of execution are used, each thread folds its
blocks of data into its own integration of the pulse profile.  On
completion of an integration, each thread places its result on a stack
where the folded results from all threads await combination.  When no
threads have data to contribute to an integration, it is processed as
in the single-threaded mode of operation.

{\bf Invalid Data Excision}:
Any data that are flagged as invalid are omitted from integration.
This feature can be disabled when studying bright pulsars and/or giant
pulses.


\subsection{Thread-safe Input Buffering}
\label{sec:buffering}

One or more of the operations in the signal processing chain may lose
samples, such that the time-bandwidth product of the output of the
operation is less than that of its input.  For example, for each block
of input data, coherent dedispersion discards at least $n_d$ samples
and will lose more if the input block size is not an integer number of
overlapping $N$-point segments.  One possible solution is to set the
raw input data block size and load overlapping blocks of raw input
data accordingly.  When only one of the operations imposes a
constraint on the raw input data block size and/or amount of raw input
data block overlap, it is trivial to set these parameters; however, it
may not be possible to satisfy the block size and overlap constraints
of more than one such operation in the signal chain.  Therefore, each
\dspsr\ operation manages its own overlap requirements using a
thread-safe input buffering scheme.  This design feature
simultaneously increases both the modularity and efficiency of the
code by reducing interdependency between signal processing components
and eliminating the need to re-process data lost by operations that
occur later in the signal path.

Each operation that loses samples simply buffers the appropriate
amount of data from the end of its current input data block to be
later prefixed to the next input data block passed to the operation.
A small amount of additional book-keeping ensures the contiguity of
consecutive input data blocks passed to each operation.  When multiple
processing threads are used, the input data blocks are distributed
across multiple signal processing paths.  Consequently, in each
thread, the blocks passed on consecutive calls to an operation are not
necessarily contiguous.  Therefore, for each component in the signal
chain that requires it, the threads share a single input buffer
through which the data to be prefixed are passed to the thread that
receives the next contiguous input block.  This method works only when
the processing threads share the same physical memory.  When threads
are run on different computational devices, such as multiple graphics
processing units (GPUs), and the cost of transmitting the buffered data
between devices is high, then \dspsr\ will resort to the raw input
data overlap strategy.


\section{Performance}
\label{sec:performance}

Digital signal processing is demanding of computational, data
communications, and data storage resources.  Therefore, the
performance of a baseband recording and processing system depends on a
number of variables, including the number of microprocessors, their
architecture and clock speed, memory size and bandwidth, cache size
and bandwidth, compiler version, and optimisation options.  To
demonstrate some of these basic considerations, a number of
illustrative benchmarks are presented using currently available
microprocessor technology.

\subsection{Central Processing Units}
\label{sec:cpus}

\begin{figure}
\includegraphics[angle=0,width=75mm]{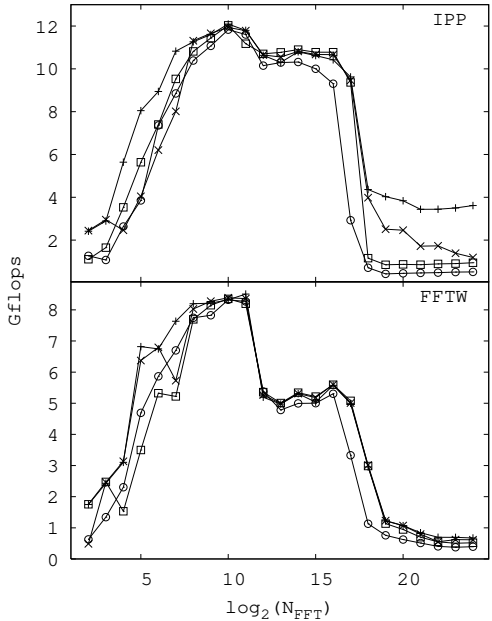}
\caption
{Fast Fourier Transform performance as a function of transform length,
\nfft.  Speed is measured in effective Gigaflops per processing core 
as described in the text.  Version 3.1.2 of the Fastest Fourier
Transform in the West (FFTW) and version 5.3.3 of the \intel\
Integrated Performance Primitives (IPP) libraries were used.  Each
panel displays the results of performing a single-precision,
complex-to-complex FFT on each of one (plus), two (cross), four
(square), and eight (circle) processing threads.  Each thread performs
a separate FFT in isolation; therefore, the aggregate speed of the
microprocessor is the value shown multiplied by the number of threads.}
\label{fig:fft_bench}
\end{figure}

For the most computationally intensive operations in \dspsr, such as
coherent dedispersion and filterbank synthesis, execution time is
dominated by computing the Fast Fourier Transform (FFT); therefore,
the rate at which data are processed is a strong function of the
performance of the FFT library.
In \Fig{fft_bench}, the speeds of two commonly used FFT libraries are
plotted as a function of transform length, \nfft, and the number of
operating threads.  Similar to the {\sc benchFFT} convention, performance
in Gigaflops is defined as
\begin{equation}
{\rm Gflops} = { 5 \nfft \log_2 (\nfft) \over t_{\mathrm{ns}} },
\label{eqn:Gflops}
\end{equation}
where $t_{\mathrm{ns}}$ is the average time required to perform a
single FFT in nanoseconds.  
These benchmarks were performed on a workstation with dual \intel\
\xeon\ E5345 (Clovertown) processors, each with 4 cores running at
2.33\,GHz, 4 $\times$ 64\,kB of level 1 (L1) cache, and 2 $\times$
4\,MB of level 2 (L2) cache,
%
%
connected via 10.66\,GB/s bus to 16 GB of Random Access Memory (RAM).

\Fig{fft_bench} demonstrates the impact of L1 and L2 cache on FFT 
performance.  FFT speed increases with greater FFT length until the
32\,kB of L1 cache reserved for data is filled (note that
$\nfft=2^{12}$ = 4\,k corresponds to 64\,kB of memory required for input
and output arrays).  At this point the speed drops (most
notably for FFTW) and remains constant before dropping again when the
L2 cache is filled ($\nfft=2^{18}$ = 256\,k corresponds to 4\,MB of
memory).  As each of the 8 processing cores is equipped with its
own L1 cache, there are no significant differences in FFT performance
as a function of the number of operating threads for $\nfft < 2^{17}$.
However, because each L2 cache is shared by two processing cores, the
effective L2 cache per core is halved when 8 processing threads are
used; in this case, as seen in \Fig{fft_bench}, performance drops when
$\nfft=2^{17}$=128\,k.

Using the performance measurements plotted in \Fig{fft_bench}, \dspsr\
can select both the optimal transform length and the best FFT library
to employ during phase-coherent dispersion removal and synthetic
filterbank formation.  The selection is made by eliminating the
$\Offt(N)\simeq5N\log_2N$ assumption and using the measured time
required to perform an $N$-point FFT while minimising \Eqn{flops} or
\Eqn{filterbank_flops}.  Particularly for shorter transform lengths,
performance is significantly improved by this optimisation.

\begin{figure}
\includegraphics[angle=0,width=80mm]{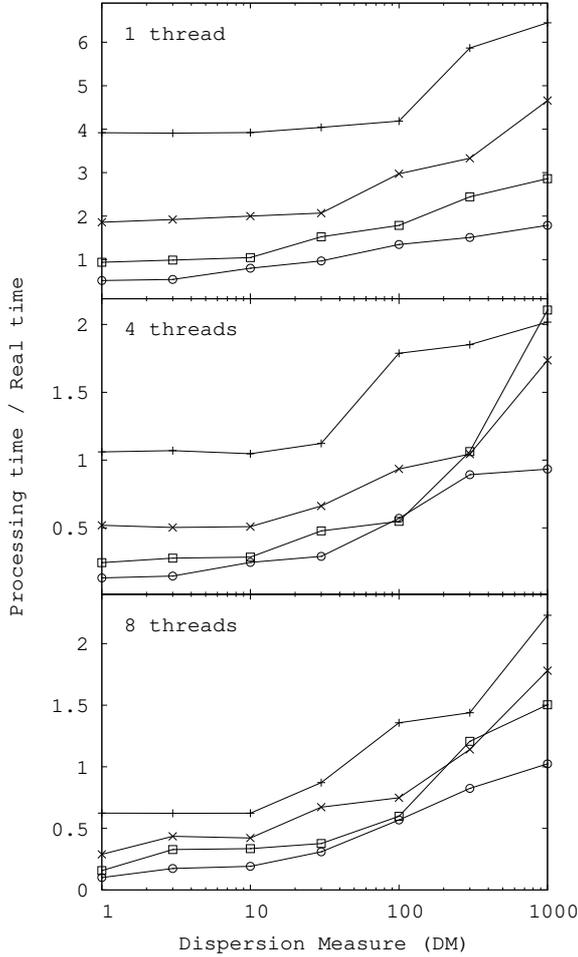}
\caption
{Processing time to real time ratio on dual \intel\ \xeon\ E5345
processors as a function of dispersion measure $DM$, centre frequency
$\nu$, and bandwidth $\delta\nu$.  The benchmark was performed using
one (top), four (middle), and eight (bottom) processing threads.  Each
panel plots four instrumental configurations, including
$\nu=3000$\,MHz with $\delta\nu=64$\,MHz (plus), 
$\nu=1500$\,MHz with $\delta\nu=32$\,MHz (cross),
$\nu=750$\,MHz with $\delta\nu=16$\,MHz (square), and
$\nu=375$\,MHz with $\delta\nu=8$\,MHz (circle).
Processing time is the total time required to perform all steps
in the typical pulsar signal processing path described in the text.
Fourier transforms were performed using the optimal transform length
and the best of either the FFTW or IPP libraries.}
\label{fig:dm_bench}
\end{figure}

Overall performance is evaluated by measuring the time required to
completely process a block of data normalised by the real time spanned
by that block of data (the real time is equal to the sampling interval
times the number of samples in the block).  \Fig{dm_bench} plots the
processing time to real time ratio as a function of dispersion measure
for four different instrumental configurations.  For each
configuration, the processing time is that required to convert 8-bit
digitised data to single-precision floating-point complex numbers,
form a filterbank with 500 kHz resolution while performing
phase-coherent dispersion removal, detect the polarisation of the
signal, and fold all frequency channels and polarisation parameters
with a resolution of 1024 phase bins.

\begin{figure}
\includegraphics[angle=0,width=80mm]{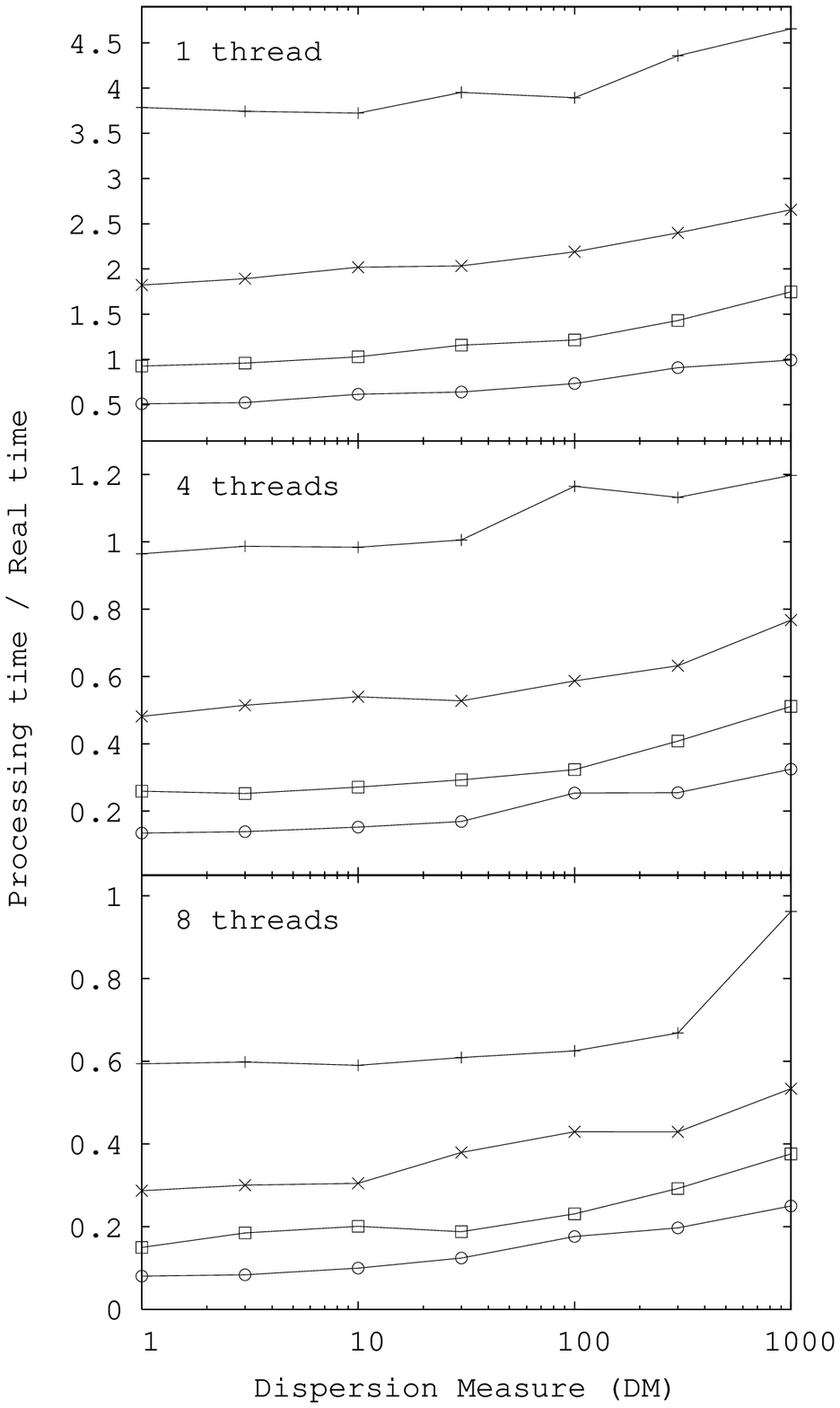}
\caption
{Processing time to real time ratio on dual \intel\ \xeon\ E5520
processors, as in \Fig{dm_bench}.}
\label{fig:dm_bench_Nehalem}
\end{figure}

%
%
At low $DM$ in \Fig{dm_bench}, where the FFT length is smaller than
the L2 cache, \dspsr\ performance scales roughly linearly with the
number of processing threads.  At the largest $DM$, 8 threads perform
slightly worse than 4 threads owing to competition for L2 cache (on
the Clovertown, each L2 cache is shared by two processing cores).
To further illustrate the importance of cache size and memory
bandwidth, the same benchmark was repeated on a workstation with dual
\intel\ \xeon\ E5520 (Nehalem-EP) processors, each with 4 cores running
at 2.26\,GHz, 4 $\times$ 64\,kB of L1 cache, 4 $\times$ 256\,kB of
mid-level (L2) cache,
%
%
and 8\,MB of L3 cache, connected via 25.6\,GB/s bus to 24\,GB
of RAM.  The results are plotted in \Fig{dm_bench_Nehalem}.
Performance at the lowest DM is not significantly improved, indicating
that the problem is bound by the processor speed in this regime.  At
the highest DM, where the FFT no longer fits in L2 cache, data are
processed as much as 4 times faster on the E5520, indicating that
memory bandwidth must be the limiting factor on the E5345.

The relatively modest bandwidths used for the performance benchmarks
presented in \Figs{dm_bench}{dm_bench_Nehalem} are roughly an order of
magnitude smaller than currently available with modern pulsar
instrumentation.  To process larger bandwidths in real time, the
digitised signal must be divided between multiple workstations, the
number of which depends upon the method of division.  
Using time-division demultiplexing, each machine must process the full
bandwidth for some fraction of the time.  Either the frame size must
be relatively large or the frames must overlap in time to compensate
for the data lost to edge effects, which may necessitate the use of
one or more intermediate workstations to buffer the frames.  Using
frequency-division demultiplexing, each machine must process some
fraction of the bandwidth for the full time.  To divide the signal
into frequency channels, some form of filterbank (such as a polyphase
filterbank implemented on a field-programmable gate array) must be
used.

\begin{figure}
\includegraphics[angle=0,width=80mm]{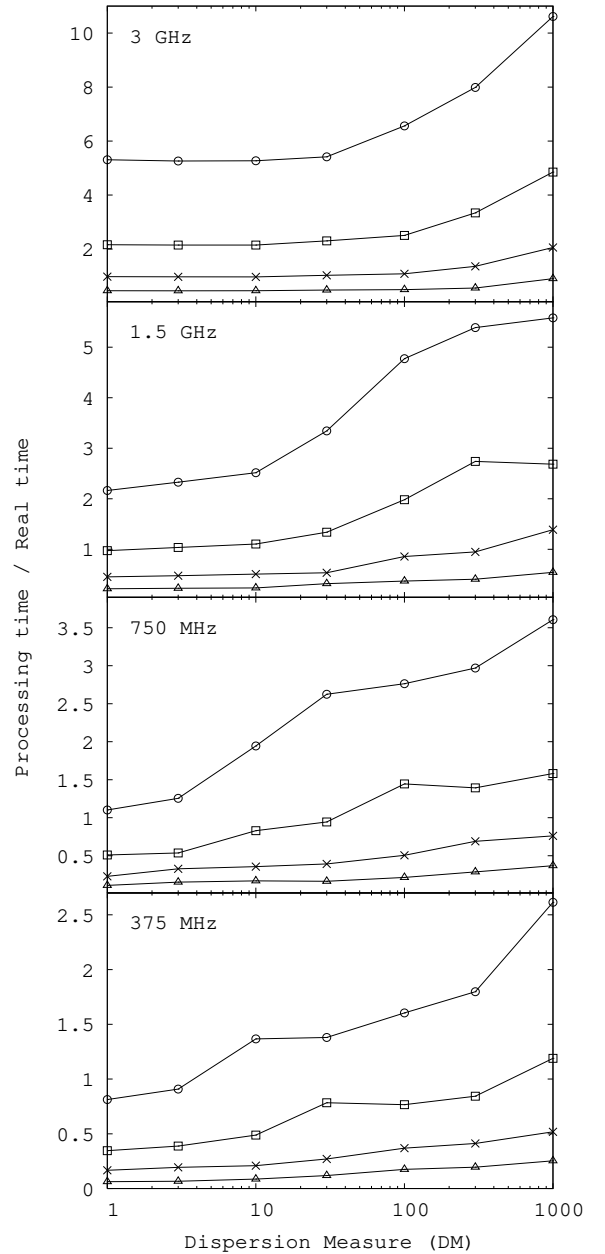}
\caption
{Processing time to real time ratio on dual \intel\ \xeon\ E5520
processors as a function of dispersion measure $DM$.  The centre
frequency $\nu$ varies from 375~MHz (bottom panel) to 3~GHz (top
panel).  In the bottom panel, the bandwidth $\delta\nu$ varies as
 $\delta\nu=8$\,MHz (triangle), 
 $\delta\nu=16$\,MHz (cross),
 $\delta\nu=32$\,MHz (square), and
 $\delta\nu=64$\,MHz (circle).
Each symbol corresponds to the same relative bandwidth $\delta\nu/\nu$
in each panel.
Processing time is defined as in \Fig{dm_bench}.}

\label{fig:cpu_bench}
\end{figure}

When designing a digital signal processing software-based instrument,
the number of workstations required to process the desired bandwidth
in real time may be estimated using a benchmark similar to that
presented in \Fig{cpu_bench}.
For example, considering the top panel (3~GHz), an instrument that
uses time-division demultiplexing to process a 512~MHz band (circles)
out to a $DM$ of 1000 will require at least 11 workstations.
%
%
If the signal is first divided into four sub-bands (crosses), only 8
workstations (2 per sub-band) would be required.

\subsection{Graphics Processing Units}
\label{sec:gpus}

\begin{figure}
\includegraphics[angle=0,width=75mm]{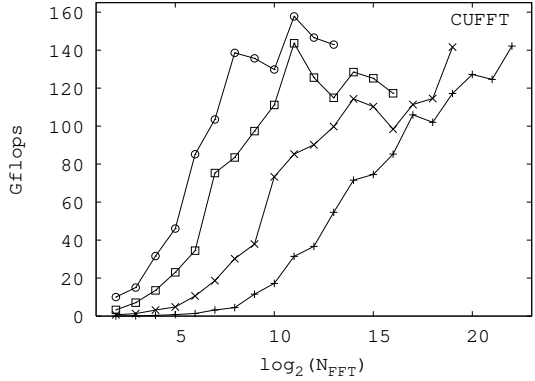}
\caption
{Filterbank performance as a function of inverse transform length
\nfft\ and number of frequency channels $N_c$.  Speed is measured in
effective Gigaflops for a $N_c=4$ (plus), $N_c=32$ (cross), $N_c=256$
(square), and $N_c=2048$ (circle) filterbank up to the limit of
$\log_2 (N_c \nfft) = 24$. Version 3.0 of the \nvidia\ \cuda\ FFT
library (CUFFT) was used.}
\label{fig:fft_cuda}
\end{figure}

On a multiprocessor device such as a graphics processing unit (GPU),
the greatest performance is achieved when all of the processors are
fully utilised.
To maximise utilisation, the \nvidia\ \cuda\ FFT library (CUFFT)
enables parallel (batched) execution of multiple transforms.
This feature is exploited by the GPU-based implementation of the
convolving filterbank algorithm to perform the $N_c$ inverse
$N^\prime$-point FFTs in parallel (refer to the algorithm at the end
of \Sec{convolving_filterbank}).
Consequently, FFT performance is a function of both $N_c$ and
$\nfft=N^\prime$, as shown in \Fig{fft_cuda}.  Here, performance in
Gigaflops is defined as
\begin{equation}
{\rm Gflops} = { 5 N_c \nfft (\log_2 N_c + 2 \log_2 \nfft) \over t_{\mathrm{ns}} },
\label{eqn:Gflops_filterbank}
\end{equation}
where the number of floating point operations is as derived in the
Appendix and $t_{\mathrm{ns}}$ is the average time in nanoseconds
required to perform the forward and backward transforms in each step
of the convolving filterbank algorithm.  These benchmarks were
performed on a workstation equipped with an \nvidia\ Fermi
architecture Tesla C2050 GPU. In contrast to the benchmarks presented
in \Fig{fft_bench}, where memory bandwidth and cache size limits
performance at large \nfft, CUFFT performs best when $N_c \times
\nfft$ is large because greater microprocessor utilisation is achieved.

Optimal GPU performance also requires minimising the amount of data
transferred between the device and host memory.  That is, once data
are on the GPU, it is best to reduce them as much as practicable on
the device.
To this end, \dspsr\ implements GPU-based detection and folding
algorithms, so that only the integrated results are transferred back
to host memory.
The folding algorithm exploits hardware-level multithreading to
maximise utilisation by launching a processing thread for each frequency
channel, polarisation, and pulse phase bin,
%
%
such that the total number of independent sums to be computed is much
greater than the number of available cores (448 on the Fermi
architecture).
By scheduling many more threads than there are processors, the latency
in thread execution incurred during retrieval of data from off-chip
memory is hidden by switching between active execution contexts (or
{\it warps}: atomic groups of threads that execute simultaneously).
Note that on the single-instruction, multiple-thread (SIMT)
architecture employed by streaming multiprocessors, separate registers
are allocated to each active thread; therefore, there is no cost
associated with switching between active warps.
If the maximum amount of data that fits into GPU memory corresponds to
less than one pulse period and the number of frequency channels is
small, then some fraction of the processors may be idle; therefore,
the current folding algorithm implementation performs best on pulsars
with short spin periods.

\begin{figure}
\includegraphics[angle=0,width=80mm]{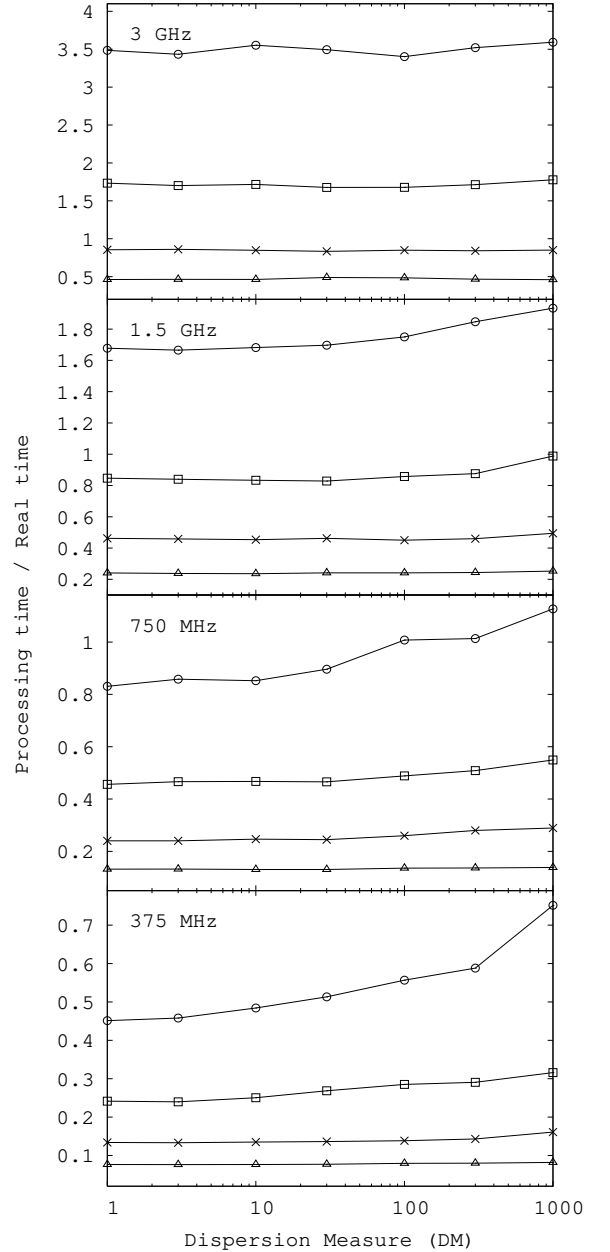}
\caption
{Processing time to real time ratio on an \nvidia\ Fermi architecture
Tesla C2050 as a function of dispersion measure $DM$, as in
\Fig{cpu_bench}.}
\label{fig:gpu_bench}
\end{figure}

The overall performance of a contemporary GPU-based instrument can be
estimated using the benchmarks presented in \Fig{gpu_bench}.  Focusing
once again on the top panel (3~GHz), to process a 512~MHz band out to
a $DM$ of 1000 in real time requires only 4 GPU-equipped workstations,
regardless of time- or frequency-division demultiplexing.  For reference,
the spin period of the pulsar used in these benchmarks is $\sim 5.76$ ms.

The software used to produce the benchmarks presented in this section
is distributed as part of the \dspsr\ package.  To facilitate
instrumental design and planning, the latest available performance
benchmarks are maintained online\footnote{http://dspsr.sourceforge.net/bench}.


\section{Conclusion}
\label{sec:conclusion}

\dspsr\ is a high-performance, general-purpose tool for radio
astronomical data reduction that has enabled a diverse range of
experiments, including 
the discovery of the giant micropulse phenomenon \citep{jvkb01}, 
a survey for sub-millisecond pulsars \citep{evb01},
high-precision pulsar timing \citep{vbb+01,hbo04,jhb+05,ojhb06,vbv+08}, 
millisecond pulsar polarimetry \citep{ovhb04},
giant pulses \citep[e.g.][]{kbm+06},
and dispersion measure variations \citep{yhc+07},
studies of giant pulses from the Crab Nebula pulsar \citep{btk08,ksv10},
the first observations of rotating radio transient polarisation \citep{khv+09},
the search for gravitational waves \citep{vbc+09,yhj+10},
and an analysis of scattering by the interstellar plasma \citep{crg+10}.
\dspsr\ has also been used in the analysis of giant pulse data from 
the Mileura Widefield Array Low Frequency Demonstrator
\citep[MWA-LFD;][]{bwk+07},
and early pulsar observations made with the Low Frequency Array Core
Station 1 (LOFAR CS1; B.~Stappers 2009, private communication).

Data recorded at most major radio observatories are understood by
\dspsr, which can be used to process either voltage- or power-level
signals (i.e.\ undetected or detected data).  For example,
\dspsr\ can process baseband data encoded using the VLBI Data
Interface Format (VDIF)\footnote{http://www.vlbi.org/vsi} as well as
detected data stored using the \psrfits\ file format.  The library
also contains routines to both read and write files in the format used
by {\sc sigproc}, another popular signal processing package that is
commonly used in the search for new pulsars, making \dspsr\ a useful
component in survey data reduction pipelines.
Although most of the design and testing of the software has been
completed for the analysis of radio pulsar observations, the large
majority of the algorithms may be applied to any single beam (either
single-dish or phased-array) radio observation.  For example, the
\dspsr\ library could be utilised to produce high-resolution
spectra for studies of large scale structure in extragalactic neutral
Hydrogen \citep[e.g.][]{skc+03}.

Future developments of \dspsr\ may include the adoption of industry
standards like the Open Computing Language
(OpenCL)\footnote{http://www.khronos.org/opencl} for portable GPU
programming.
GPU performance may in turn facilitate the computation of
new statistical quantities.  For example, the cyclostationary
statistics of the electromagnetic field promises new insight into the
phenomenon of diffractive scintillation (P.~Demorest 2008, private
communication) and may enable the inversion of multipath scattering 
effects that arise in the interstellar medium \citep{wksv08}.
Although formation of the cyclic power spectrum is computationally
prohibitive, it may be feasible to regularly observe this statistic
using GPU technology.

\dspsr\ provides the capacity for real-time phase-coherent dispersion 
removal over a comprehensive range of observing configurations,
including those planned for the upcoming generation of radio
telescopes such as the Australian Square Kilometer Array Pathfinder
\citep[ASKAP;][]{jtb+08} and the Karoo Array Telescope
\citep[MeerKAT;][]{jon09}.
The ASKAP Boolardy Engineering Test Array (BETA) will operate between
700 MHz and 1.8 GHz and provide a $2\times300$ MHz band.
%
%
The MeerKAT Precursor Array (MPA -- also known as KAT-7) will provide
a $2\times256$ MHz band within the frequency range from 1.2 to 1.95
GHz.
%
%
Benchmarks demonstrate that \dspsr\ can comfortably process these
signals in real time using only 4 workstations, each equipped
with a single GPU.
Considering its exemplary performance, extensive functionality, and
degree of maturity, both scientists and engineers are encouraged to
take advantage of this well-tested open-source digital signal
processing library for radio astronomy.

\section*{Acknowledgments}

Throughout the development of this software, we have benefited from
the advice and input of many colleagues, especially Dan Stinebring,
Josh Kempner, Matthew Britton, Stuart Anderson, Russell Edwards,
Haydon Knight, Ben Stappers, Ben Barsdell and Jonathon Kocz.  We are
grateful to those who have implemented support for the interpretation
of data from various instruments, including Paul Demorest, Aidan
Hotan, Mike Keith, Jonathan Khoo, Karuppusamy Ramesh, Jayanta Roy,
Kevin Stovall, and Craig West.  We thank Xenon Systems and the
Victorian Partnership for Advanced Computing (VPAC) for provision of
the \nvidia\ Fermi GPU and assistance with its use.  We acknowledge
\intel\ for support of our Square Kilometre Array research programme.
The Parkes Observatory is part of the Australia Telescope which is
funded by the Commonwealth of Australia for operation as a National
Facility managed by CSIRO.


\appendix

\section{Floating Point Operations}
\label{app:flops}

The number of floating point operations required to create a synthetic
filterbank and perform phase-coherent dispersion removal is a function
of the number of frequency channels produced by the synthetic
filterbank, $N_c$, and the size of the dispersion response function in
the output frequency channel with the lowest centre frequency,
$N^\prime$.  This is true for both real-valued and complex-valued
input data.  Although twice as many real-valued input time samples are
required to obtain the same resolution in the frequency domain, the
$2N$-point real-to-complex Fast Fourier Transform (FFT) is readily
computed with roughly the same efficiency as the $N$-point
complex-to-complex FFT, as described in Section~12.3 of Numerical
Recipes \citep{ptvf92}.

In the case of the deprecated filterbank, the forward $N_c$-point FFT
is repeated $N^\prime$ times, requiring 
\[
N^\prime \times 5 N_c \log_2 N_c
\]
operations.  Convolution is then performed independently on each of
the $N_c$ channels using a forward and backward $N^\prime$-point FFT,
costing
\[
2 N_c \times 5 N^\prime \log_2 N^\prime
\]
operations.
In the case of the convolving filterbank, the single $N_c N^\prime$ forward
FFT requires
\[
5 N_cN^\prime \log_2 N_cN^\prime
\]
operations.  For each of the $N_c$ channels, a backward $N^\prime$-point
FFT is performed, costing
\[
N_c \times 5 N^\prime \log_2 N^\prime
\]
operations.  Although both methods require a total of
\[
5 N_c N^\prime (\log_2 N_c + 2 \log_2 N^\prime)
\]
floating point operations, the speed with which those operations are
performed will depend on the implementation details of the FFT and the
architecture of the computing device.  As shown in \Fig{fft_bench},
FFT libraries typically achieve greater efficiency as the transform
size is increased until the problem no longer fits in L1 cache.
Therefore, if $N_c$ is smaller than the L1 cache-limited transform
length, then the convolving filterbank may provide better performance than
the deprecated method.  However, if the product of $N_c \times
N^\prime$ is large, such that the memory required to compute the FFT
is larger than the L2 cache size, then the deprecated method may
provide better performance.  Regardless of computational performance
considerations, in light of the spectral leakage artifacts introduced
by the deprecated filterbank method (see \Fig{spectral_leakage}), use
of the convolving filterbank is recommended.


\end{document}